\documentclass{aa}
\usepackage{graphicx}
\usepackage{txfonts}
\usepackage{natbib}
\usepackage{xcolor}
\usepackage{hyperref}

\begin{document}

   \title{Continuum polarization reverberation mapping of AGNs}
   
   \author{P.~A.~Rojas~Lobos
   \and R.~W.~Goosmann
   \and J.~M.~Hameury
   \and F. Marin}
   
   \authorrunning{Rojas Lobos et al.}

   \institute{Universit\'e de Strasbourg, CNRS, Observatoire astronomique de Strasbourg, UMR 7550, F-67000 Strasbourg, France \\
   email: jean-marie.hameury@astro.unistra.fr}

   \date{Received 25 February 2020; Accepted 8 April 2020}

  \abstract
  {The determination of the size and geometry of the broad line region (BLR) in active galactic nuclei is one of the major ingredients for determining the mass of the accreting black hole. This can be done by determining the delay between the optical continuum and the flux reprocessed by the BLR, in particular via the emission lines.}
  {We propose here that the delay between polarized and unpolarized light can also be used in much the same way to constrain the size of the BLR; we check that meaningful results can be expected from observations using this technique.} 
  {We use our code {\sc Stokes} for performing polarized radiative transfer simulations. We determine the response of the central source environment (broad line region, dust torus, polar wind) to fluctuations of the central source that are randomly generated; we then calculate the cross correlation between the simulated polarized flux and the total flux to estimate the time delay that would be provided by observations using the same method.}
  {We find that the broad line region is the main contributor to the delay between the polarized flux and the total flux; this delay is independent on the observation wavelength.}  
  {This validates the use of polarized radiation in the optical/UV band to estimate the geometrical properties of the broad line region in type I AGNs, in which the viewing angle is close to pole-on and the BLR is not obscured by the dust torus.}  


\maketitle

%________________________________________________________________
\section{Introduction}
\label{Introduction}

Since the discovery of broad emission lines in the polarized spectrum of NGC~1068 \citep{ma83,am85}, and in 3C~234 \citep{a84b}, a radio-loud AGN, our current view of an active galactic nucleus (AGN) has been that of an accreting black hole surrounded by its accretion disc, and, lying close to the equatorial plane, in increasing distance order, the broad line region (BLR) and a dust torus \citep{a93}. At much larger distances, of the order of hundreds of parsecs, the narrow line region (NLR) is found, and can be spatially resolved. Outflows, in form of a polar wind (PW) and/or a jet may also form and connect the central engine to the NLR. Differences in the viewing angle of these systems account for most of the differences in their observational appearance: when the AGN is seen close to edge on, the broad line region is hidden by the dust torus and broad lines are normally not observed; the AGN is of type 2. When on the other hand the inclination is such that the BLR is directly visible, the AGN is of type 1. 

Until very recently, except for the NLR and the jet at large distances, none of these regions could be spatially resolved, and our knowledge on their structure was therefore indirect. Recent interferometric observations of the quasar 3C~273 \citep{gsd18} and of Sgr~A$^*$ \citep{gaa18} have indeed been able to provide spatial information on the innermost parts of AGNs -- the BLR in the case of 3C~273, and regions at a few Schwartzschild radii in the case of Sgr~A$^*$. A few other sources have been observed using interferometry techniques; first NGC~1068 with the MiD-Infrared Interferometric Instrument {\it MIDI} on the Very large Telescope Interferometer (VLTI) \citep{jmr04,rjr09} or with {\it ALMA} \citep{gcr16}, and, later, a few other ones with {\it Gravity} \citep{gds19} that have partially resolved the hot dust regions. These observations, however, are often limited to visibility measurements for nearby sources. The Event Horizon Telescope has provided images of the immediate vicinity of the central back hole of M~87 \citep{eht19}, but these cannot constrain the more distant regions, such as the external parts of the accretion disk or the BLR. Despite these spectacular progresses, adding new observational tools is therefore most important; polarimetry has proven in the recent years that it can be very efficient for this purpose \citep[see for example the detection of broad lines in the polarized spectrum of NGC~1068 by][a strong evidence that type I and type 2 AGNs form a unified class]{am85}  because it traces photons that have scattered in the environment of the central source.

The determination of the mass of the central supermassive black hole in active galactic nuclei is a key ingredient for understanding the structure, formation and evolution of these objects, and more generally of galaxies themselves. Several techniques can be used, among which reverberation mapping is popular because of its simplicity. In this technique, the motion of the BLR clouds is assumed to be virialized; their velocity can be estimated from the line velocities, and their distance to the central black hole is determined by the delay between variations of the continuum flux and variations of the line intensity \citep[see e.g.][]{bm88,p93,zkp11}. Delays of a few days are usually observed \citep{lkn18}, corresponding to BLR sizes a few light-days, but longer delays have been found in a number of cases, such as CTS~252 \citep[190 days in the quasar rest frame,][]{lkn18}, and CTS~C30.10 \citep[about 560 days in the rest frame,][]{cor19}. The mass of the central object directly follows from Newton's law. 

In this paper, we consider another possible way to constrain the size of the BLR, namely the use of the delay between variations of the polarized flux and of the continuum flux. This method has been proposed in \citet{sgg05} and later used by \citet{ggm12} who found that the polarized flux in the B band in NGC~4151 follows the total flux in the B band with a delay of $8 \pm 3$ days. However, polarization of the optical light in AGN arises from scattering of the continuum emitted by the central source on the environment of the black hole, that includes the BLR, but also other circumnuclear regions such as the polar wind, and the dust torus. In the case of the BLR and the polar wind, scattering is usually considered to be due to electrons, whereas in the case of the torus, scattering is due to dust. In some models, dust scattering plays an important role in the BLR \citep[see e.g.][]{ch11,mg13,clh17}; these are not considered here. One therefore needs to check that the presence of these outer regions does not significantly affect the conclusion one might draw from the determination of this lag. One also needs to relate in a more precise way the measured time-lag to the properties of the BLR.

In a previous paper \citep[][hereafter Paper I]{rgm18}, we calculated the average time delay of polarised photons that underwent one or several scatterings as compared to photons propagating from the central source without having suffered any interaction. This mean delay is, however, impossible to determine directly from observations. This is because it includes the contribution of scattering by distant regions that would be very difficult to observe since the propagation time might be longer than the observation period over which luminosity variations are observed, and also because the ability to detect time lags depends on the coherence properties of the central source signal. The very notion of "average" delay is also poorly defined; it can for example refer to the standard mean as well as to the median. These two quantities may differ very significantly as we shall see here.

In this paper, we address the question of the comparison to observations of the predicted time-lag for scattered photons. We simulate observations of an AGN seen at low inclination, assuming that the central source varies randomly. In Section 2, we convolve the source luminosity with the transfer function that we obtain using polarized transfer simulations form our STOKES code, initially developed by \citet{gg07} and further upgraded by \citet{mgg12,mgg15,m18} in order to determine the predicted polarized and unpolarized fluxes. We do not include the contribution of the host galaxy that dilutes the light from the central source \citep{m18}, but because the starlight contribution is not expected to vary on short time scales, this should not affect our results. We note, however, that correcting for the starlight contribution is crucial for determining the slope of the BLR radius-luminosity relation \citep{bdg13}. We then calculate the cross-correlation of the polarized and total flux, and we present our results in Section 3. We show that the cross-correlation of the polarized and total flux is mainly sensitive  to scattering in the BLR, and, provided that the optical depth of the BLR is not too large, provides a measurement of its outer radius. We stress out that our aim is not to reproduce the 8 days delay observed by \citet{ggm12} in NGC~4151, but rather to test the method and see how the predicted time-lag depend on the geometry and on the parameters of the system.

\section{Model}
\label{Model}

\begin{table}
\caption{Parameters of the scattering regions}
\label{tab:param}
\centering
\begin{tabular}{l c c c c }     
\hline
component                & torus         & flared disk  & BLR             & polar wind \\
\hline
$r_{\rm in}$  (pc)       & 0.061         & 0.061        & 0.0067          & 0.0067 \\
$r_{\rm out}$ (pc)       &15.061         &15.061        & 0.0577          & 30.0067 \\ 
$\theta$ (degrees)       & 60            & 60           & 60              & 30 \\         
$\tau_{\rm opt}$         & 150           & 150          & 1               & 0.3 \\
composition              & dust          & dust         & electrons       & electrons \\
\hline 
\end{tabular}
\end{table}

\begin{figure}
\includegraphics[width=\columnwidth]{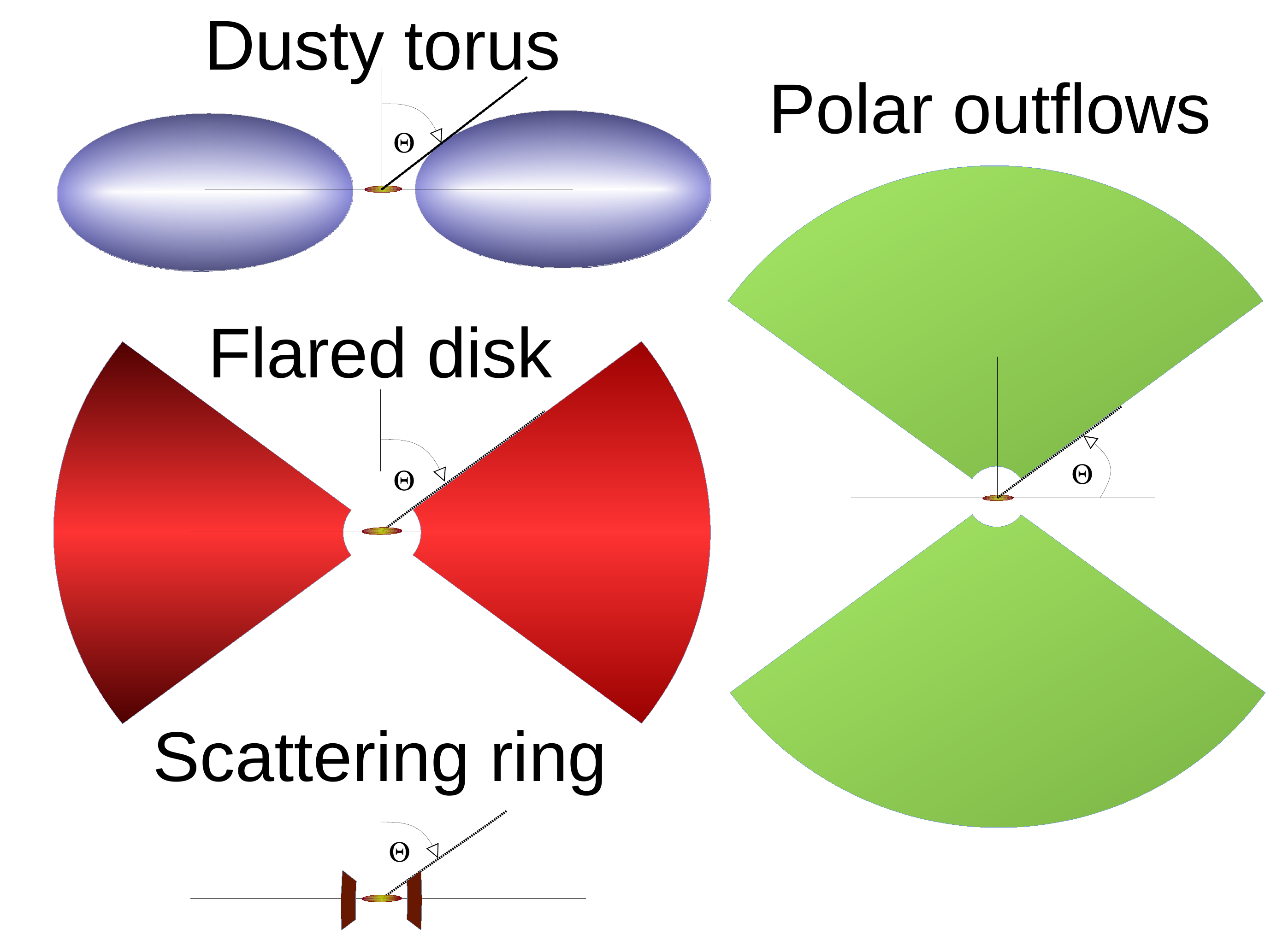}
\caption{Geometry. Left top: classical torus geometry (blue) with elliptical cross section. Left middle: the extended flared-disk geometry (red) with the wedge-shaped cross section. Left bottom: Scattering ring (brown) representing the BLR. Right: polar outflows (green). Figures from \citet{rgm18}.}
  \label{Fig:geometry}%
\end{figure}

Figure \ref{Fig:geometry} shows the geometry we use here, that is the same as in \citet{rgm18}, namely a central point source surrounded by the BLR modelled as a scattering ring with inner and outer ring radius $r_{\rm s,in}$ and $r_{\rm s,out}$ respectively; this ring has a conical cross section with half opening angle $\theta_{\rm s}$. This equatorial scattering region is requested to account for the polarization properties of Seyfert 2 galaxies \citep{sra04}; its optical depth must not be too small if this ring is to play any role at all, and it cannot be too large either. \citet{sra04} consider a ring with a Thomson optical depth of 1. The dusty absorber is assumed either to have the same shape with parameters $r_{\rm t,in}$, $r_{\rm t,out}$ and $\theta_{\rm t}$, or to have a doughnut shape. The dimensions we consider here are the same as in Paper I, and are given in Table \ref{tab:param}. The dust torus and flared disk can be either homogeneous or clumpy; in the latter case, the filling factor is 25\% and the cloud radius is 0.2~pc in the torus case, and 0.6~pc in the flared disk case. Outflows in AGNs are optically thin, with $\log N_{\rm H}$ usually in the range 20 -- 22 for warm absorbers and 22 -- 24 for ultrafast or non-ultrafast outflows \citep{tcr13}. We consider here a polar wind with total radial optical depths of 0.3, clearly at the upper end of the allowed range, in order to maximize its effects.

In the following, the inclination $\Phi$ of the AGN is taken to be 30 degrees ($\Phi = 0$ for a face-on AGN), i.e. the viewing angle corresponds to a type 1 AGN.

\subsection{Source term}

The response of the environment of the central source can be written as:
\begin{equation}
    L(\lambda,t) = \int_{0}^{\infty} \Psi(\lambda,\tau) S(t-\tau,\lambda) d\tau
    \label{eq:tfsrm}
\end{equation}
where $L$ is the observed luminosity, $S$ is the time-dependent luminosity of the central source, and $\Psi$ is the transfer function, i.e. the response of the system to an impulsive source  luminosity, $S(t) = \delta (t)$. All quantities depend on the observation wavelength $\lambda$. We consider three possible values of $\lambda$, corresponding to the visible (5500 \AA), blue (4500 \AA) and ultraviolet (2500 \AA) domains. $\Psi$ is calculated using our STOKES code that has been modified in order to calculate the path difference between scattered photons and photons that reach directly the observer without having suffered interactions.

The time variability of AGNs is often described by a damped random walk \citep{kbs09,zkk13}, in which a correcting term is added to a random walk model to push the deviations back to their mean value for time scales larger than a characteristic time scale of the order of months to years, depending on the black hole mass \citep{mik10}. For time scales shorter than this characteristic time scales, both processes give similar light curves, and the power spectrum density (PSD) is a power law with index $-2$. For very short time scales, {\it Kepler} observations of AGNs \citep{aku18} have shown that the PSD at high frequencies ($10^{-6} - 10^{-4}$ Hz, corresponding to time scales of a few hours to a couple of days) is still a power law, but with an index varying from sources to sources, usually of the order of $-2$ to $-3$. More general methods exist for simulating a light curve, such as the one proposed by \citet{tk95} for generating a signal with a power law PSD with a given index, or the continuous-time autoregressive moving average (CARMA) model \citep{kbs14}, in which the PSD can be expressed as a sum of Lorentzian functions, and is therefore extremely flexible and able to model a broad range of PSDs.

Because we are interested in time scales of the order of days to weeks, and in order to avoid introducing additional free parameters, we generated the source light curve according to a random walk scheme, i.e.
\begin{align}
& t_i = i \Delta t \\ 
& S(t_i) = S(t_{i-1}) + 2 \beta (r_i-1/2)
\label{eq:sterm}
\end{align}
where $r_i$ is a random number uniformly distributed in the range $[0-1]$, $\beta$ is a constant that determines the variability of the source light curve, taken here to be $\beta = 0.07$, and $\Delta t$ is the time interval between two discrete source points, taken here to be 0.60 days. We checked numerically that the PSD of the light curve generated according to this process was indeed a power law with index $-2$. The value of $\beta$ determines the rms variation of the source observed on a time scale $\Delta t$; it varies from source to source and depends on the observation wavelength as well as on the degree of dilution of the AGN light by the host galaxy. As we shall see later (see Fig. \ref{Fig:lc}), a $\beta$ of 0.07 results in variations by factors of two on time scales of a few years, in agreement with what is observed in e.g. NGC~4151 \citep{ggm12}. Our method therefore completely omits the red part of the PSD, as well as the high frequency component. We discuss in Sect. \ref{sec:results} the influence of the statistical model used to describe the source term on the delays one can obtain, and we show that our results are robust and do not depend significantly on the assumptions made on the time variability of the source.

\begin{figure}
\includegraphics[width=\columnwidth]{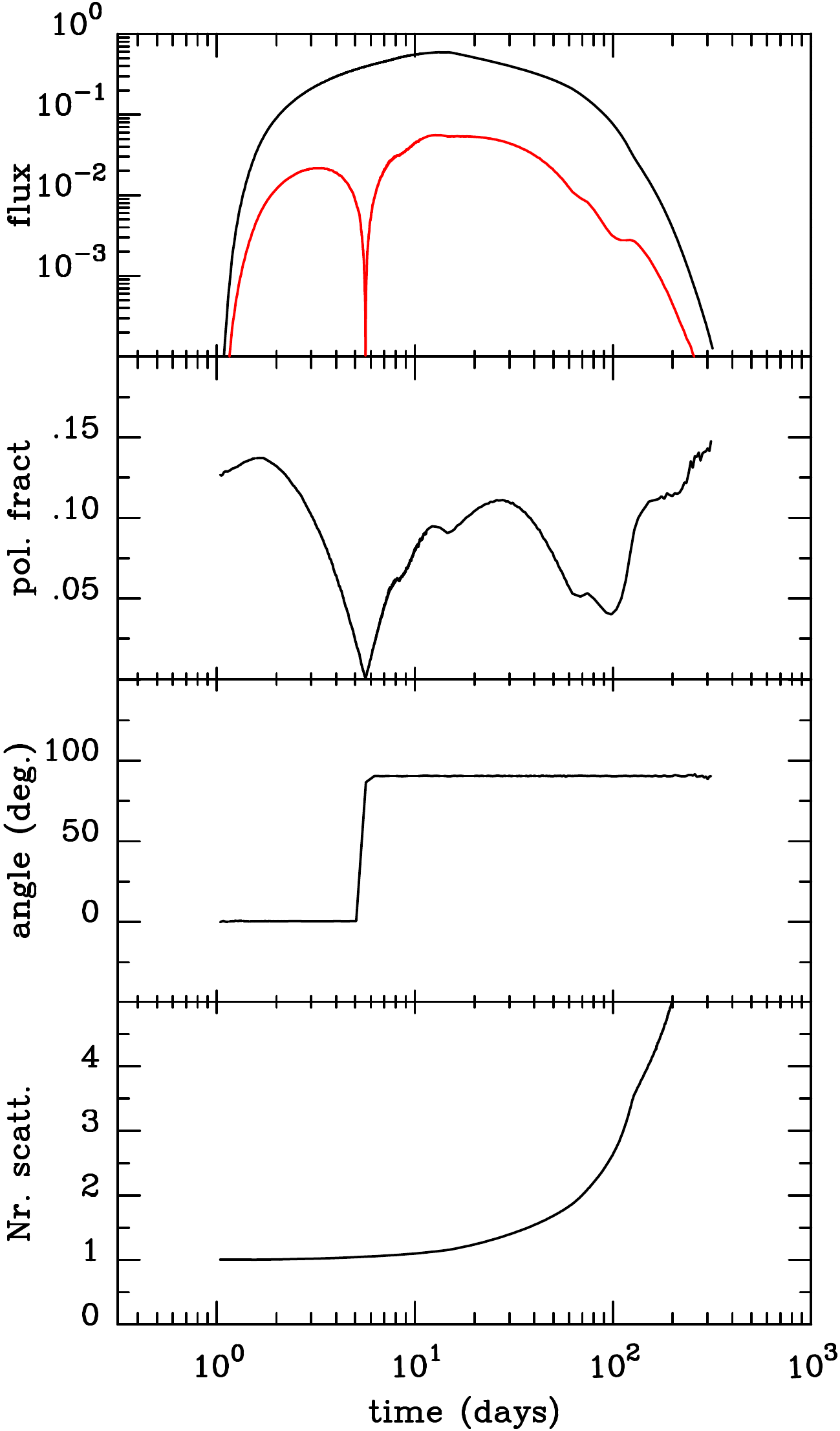}  \caption{Transfer function obtained from STOKES for equatorial ring alone. From top to bottom: total (red) and polarized (black) flux, polarization fraction, polarization angle, number of electron scattering as a function of time-lag. The viewing angle is 30$^\circ$.}
  \label{Fig:TF_ER}%
\end{figure}

\subsection{Transfer function}

\subsubsection{Equatorial ring} Figure \ref{Fig:TF_ER} shows the transfer function that we obtained considering the scattering by an equatorial ring only. The total flux is the sum of a delta function for a zero time-lag (not shown on the figure because the x-axis is logarithmic) that corresponds to photons reaching the observer without having suffered scattering, and of the scattered component than spans a broad range of time lags. There is a minimum value, due to the fact that the ring does not extend to the central black hole, given by:
\begin{equation}
\tau_{\rm min}= [1-\cos (\Phi-\theta)]  r_{\rm in}/c
\end{equation} 
For the parameters considered here, the minimum time-lag is 1.1 days, slightly smaller than the first data point on Fig.\ref{Fig:TF_ER}. The polarized flux (red line) clearly shows two maxima at 3 days and 12 days. The first maximum corresponds to the case where scattering occurs on the part of the ring that is closest to the observer, and the second maximum is due to scattering on the opposite side of the ring. This also accounts for the change in the polarization angle between the two maxima: the angle is 0 for short delays, corresponding to scattering by the fraction of the BLR that is closest to the observer, and is 90 degrees for longer delays, corresponding to a situation where the full inner BLR contributes. The position of the first maximum decreases with increasing viewing angles, and disappears when the viewing angle reaches 55$^\circ$, while the position of the second and main maximum does not depend significantly on the viewing angle. The polarized fraction is quite significant, of the order of 10 to 15\%, as expected when the scattering angle is close to $\pi/2$ \citep{c60}, and is slightly larger for the first than for the second maximum. For time delays of up to a few months, the average number of scatterings of polarized photons is very close to unity, because the optical depth of the ring is moderate: we consider an optical depth of 1 here. Long time delays correspond to multiple scatterings and have a low probability, hence the decrease in both the total and polarized flux. One should also note that the maxima of the transfer function are relatively broad.

The time-lag corresponding to an "average" scattering photon should correspond to a maximum of $\tau \Psi(\tau)$ that is proportional to the probability that the time lag is in the range $[\tau,\tau(1+\epsilon)]$ where $\epsilon$ is some fixed quantity. In the case shown in Fig. \ref{Fig:TF_ER}, this maximum occurs for a time-lag of 33 days, larger by a factor 2 than the position of the maximum of $\Psi$. The maximum of $\tau \Psi(\tau)$ corresponds quite well to the median time lag $\tau_{\rm m}$ defined such as an equal number of photons suffer a lag smaller or larger than $\tau_{\rm}$, and that is equal to 31 days. It is smaller than the mean time lag, defined as:
\begin{equation}
\label{eq:tau_av}
<\tau> = \frac{\int_0^\infty \tau \Psi(\tau) d\tau}{\int_0^\infty \Psi(\tau) d\tau}
\end{equation}
and that is equal to 43 days in the case considered here. The difference between $\tau_{\rm m}$ and $<\tau>$ is due to the long tail of $\Psi$.

This "average" scattering photon interacts inside the scattering ring at a distance of typically $r_{\rm out}/2$ to $r_{\rm out}/3$,because the optical depth is unity; this distance is 0.02 -- 0.03~pc, and the corresponding propagation time is 24 to 36 days, accounting for a geometrical factor that is slightly smaller than unity for the inclination considered here. There exists a secondary maximum, corresponding to the first peak seen in Fig. \ref{Fig:TF_ER}, that is smaller by a factor 15 than the main peak of $\tau \Psi(\tau)$, and that therefore plays no major role.

\subsubsection{Dust torus} 
\begin{figure}
\includegraphics[width=\columnwidth]{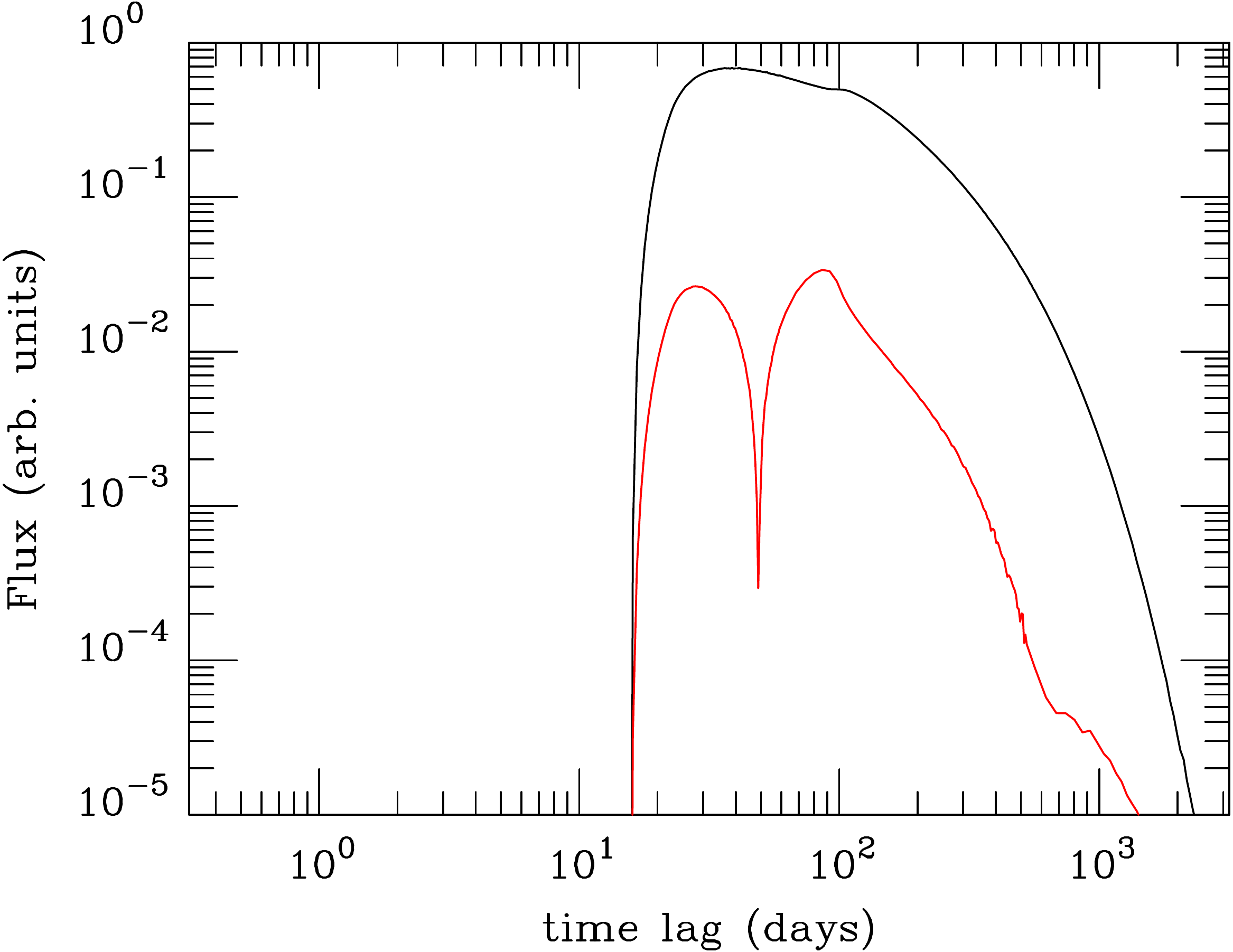}
  \caption{Transfer function obtained from STOKES for a uniform dusty torus alone. The total (black) and polarized (red) fluxes are shown.}
  \label{Fig:TF_UDT}%
\end{figure}

In the case of a dusty torus, time delays are much longer. Figure \ref{Fig:TF_UDT} shows the transfer function in the case where the torus is uniform. The two maxima correspond, as for the equatorial ring, to the inner parts of the torus that are closest (resp. farthest) to the observer. The torus is optically thick, and one expects that the "average" scattered photon has interacted with the circumnuclear medium in regions where the optical depth is of order of unity. Given the geometrical shape of the torus, these regions are mostly located close to the inner edge of the torus. More precisely, in the case of a uniform medium, the radial optical depth at a point located at a distance $r$ from the central black hole is $(r - r_{\rm in})/(r_{\rm out}-r_{\rm in}) \tau_{\rm tot}$ where $\tau_{\rm tot}$ is the total optical depth of the torus. An optical depth of 2/3 is reached at a distance $r=r_{\rm in} + r_{\rm out}/\tau_{\rm tot}$ since $r_{\rm out}$ is much larger than $r_{\rm in}$ For the parameters considered here, this occurs at a distance of 0.12~pc, and hence the time-lag of an ''average' scattered photon is about 140 days, times a geometrical factor slightly smaller than unity. This compares well with the peak of $\tau \Psi(\tau )$ that occurs at $\tau$ = 92 days. The median time lag is $\tau_{m} = 98$~d, in good agreement with the position of this maximum, and is again quite smaller than the mean lag $<\tau> = 127$~d.

\begin{figure}
\includegraphics[width=\columnwidth]{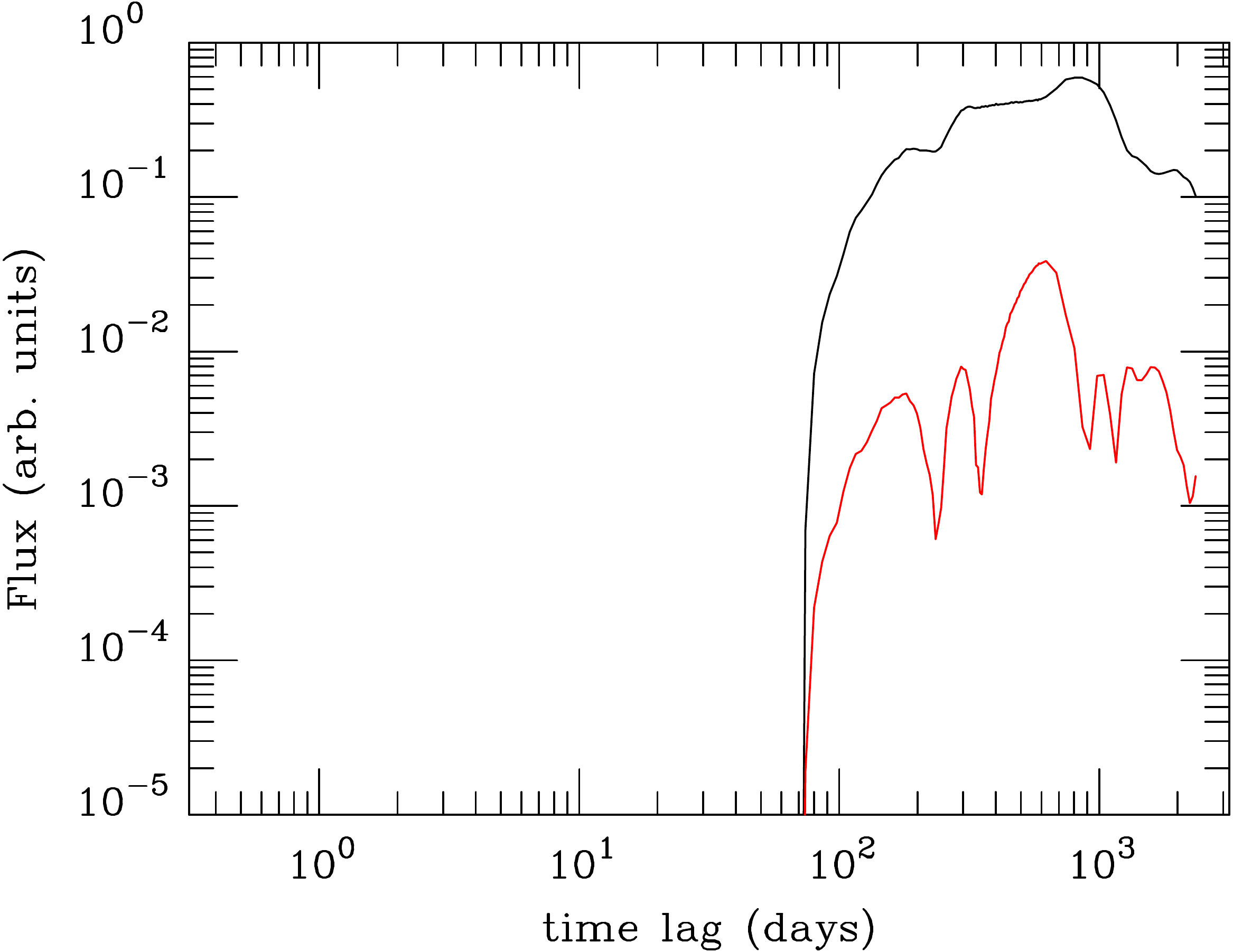}
  \caption{Same as Fig. \ref{Fig:TF_UDT} in the clumpy dusty torus case.}
  \label{Fig:TF_CDT}%
\end{figure}

\begin{figure}
\includegraphics[width=\columnwidth]{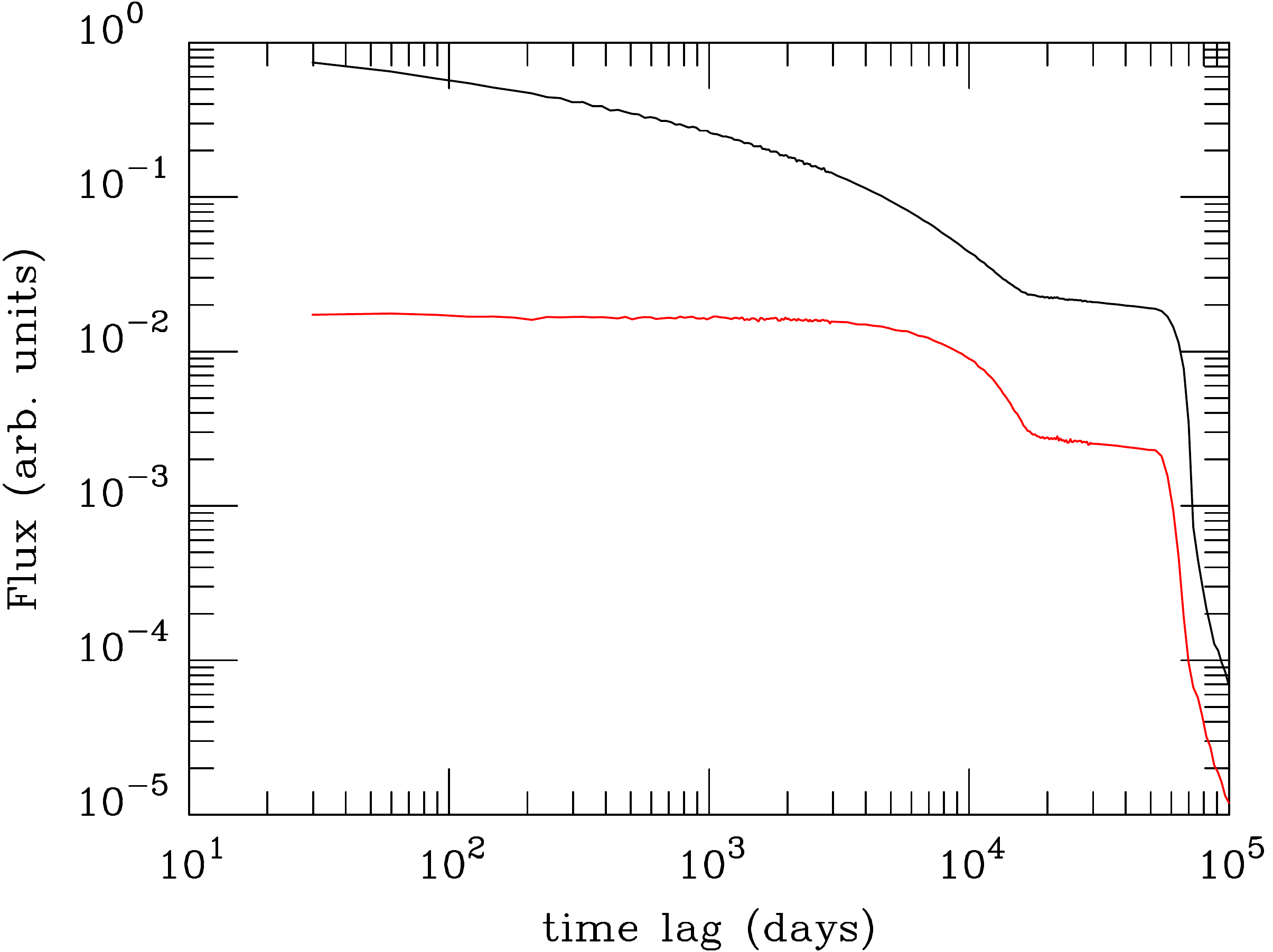}
  \caption{Same as Fig. \ref{Fig:TF_UDT} in the polar wind case.}
  \label{Fig:TF_PW}%
\end{figure}

\begin{figure}
\includegraphics[width=\columnwidth]{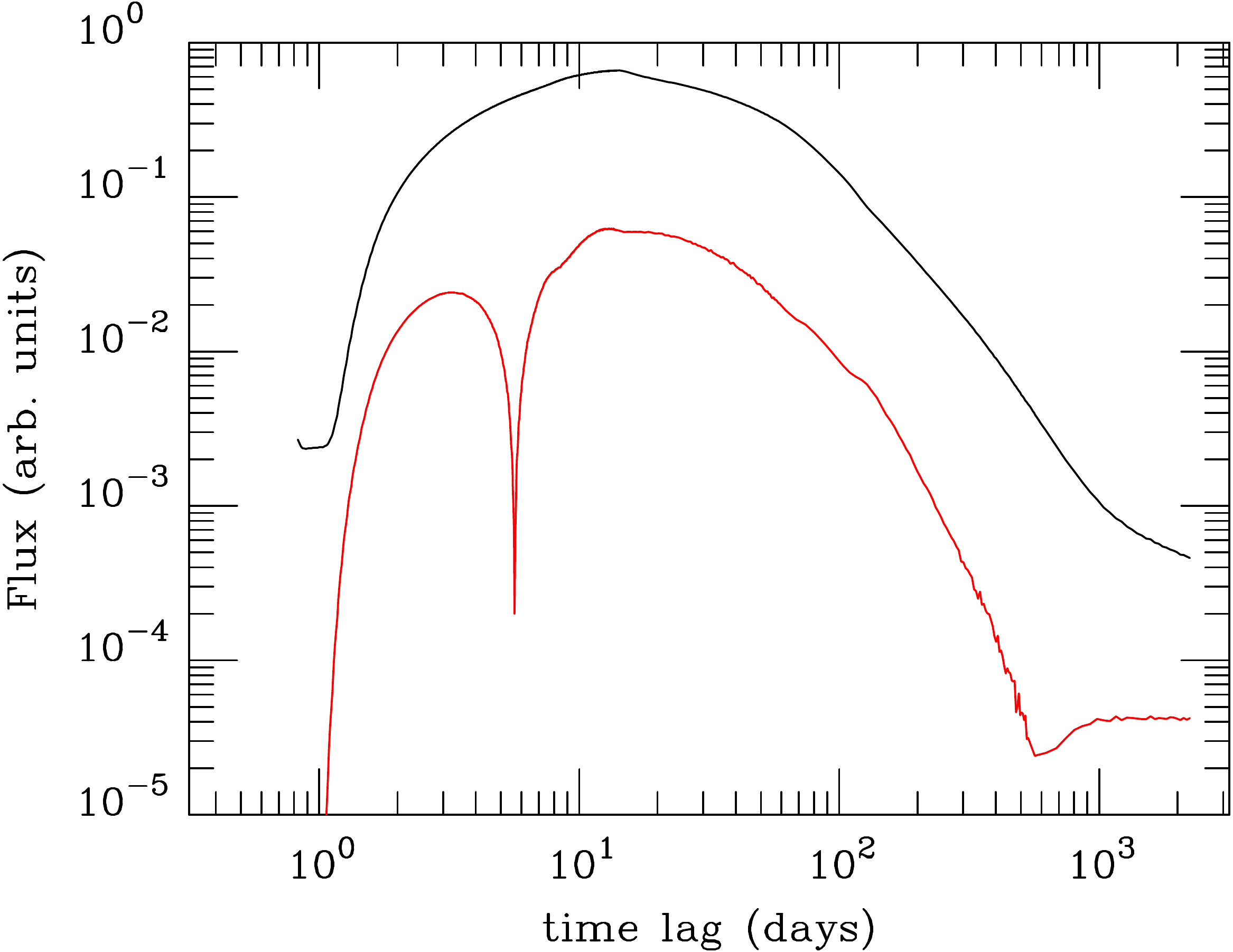}
  \caption{Same as Fig. \ref{Fig:TF_UDT} in when all components of the AGN (uniform dusty torus, polar wind and equatorial ring) are included.}
  \label{Fig:TF_full}%
\end{figure}

The situation is quite similar for the flared disk case.

Much longer delays are expected in the case of a clumpy medium because photons can penetrate much deeper than in the uniform case; the depth they can reach depends on the filling factor and on the optical depth of individual blobs. It typically varies as the filling factor to the power -2/3. However, significant deviations are expected because the cloud radius (0.2 to 0.6~pc) is larger than the inner radius of the disk. $\Psi(\tau )$ is shown in Fig. \ref{Fig:TF_CDT}, and the peak of $\tau \Psi(\tau )$ occurs at  $\tau$ = 650 days, almost an order of magnitude larger than in the uniform case. This compares well with $\tau_{\rm m} = 706$~d, and is smaller than the mean $<\tau> = 1240$~d. $\Psi$ also shows significant fluctuations that reveal the presence of individual blobs.

\subsubsection{Polar wind} Figure \ref{Fig:TF_PW} shows the transfer function for the polar wind case. The delays are long because the scattering region extends over large distances from the central source. In contrast with the previous cases, $\Psi$ is monotonic; the first plateau, for lags up to a few thousands days, corresponds to the fraction of the wind flowing towards the observer; the second one, for lags of 10,000 to about 30,000 days, is produced by the fraction of the wind flowing away from the observer. The tail for $\tau$ in the range $7 \times 10^4 - 10^5$ days is produced by photons that have scattered more than once. There are two peak of $\tau \Psi(\tau )$; the first one is is obtained for $\tau = 37$~yr, while the second one is at $\tau = 129$~yr, with a corresponding value of $\tau \Psi(\tau)$ larger than the first peak by a factor 1.15 only. The median lag is 33~yr, while the mean is 49~yr, clearly corresponding to the first of the two peaks.

\subsubsection{Full model} Finally, Fig. \ref{Fig:TF_full} shows $\Psi(\tau)$ when one includes all components of the AGN: torus, polar wind and equatorial ring. Here, the torus is assumed to be uniform. The peak in $\tau \Psi(\tau)$ is obtained for $\tau$ = 39 days; in the case of a clumpy torus, this maximum is reduced to 33 days. This happens because in the clumpy torus case, $\Psi(\tau)$ is non-vanishingly small for a much broader range of $\tau$, which tends to dilute its effect on the global curve, and also because the clumpy torus introduces delays much larger than the BLR, and has no influence on $\Psi$ for values of $\tau$ of the order of 25 days.

\begin{figure}
\includegraphics[angle=-90,width=\columnwidth]{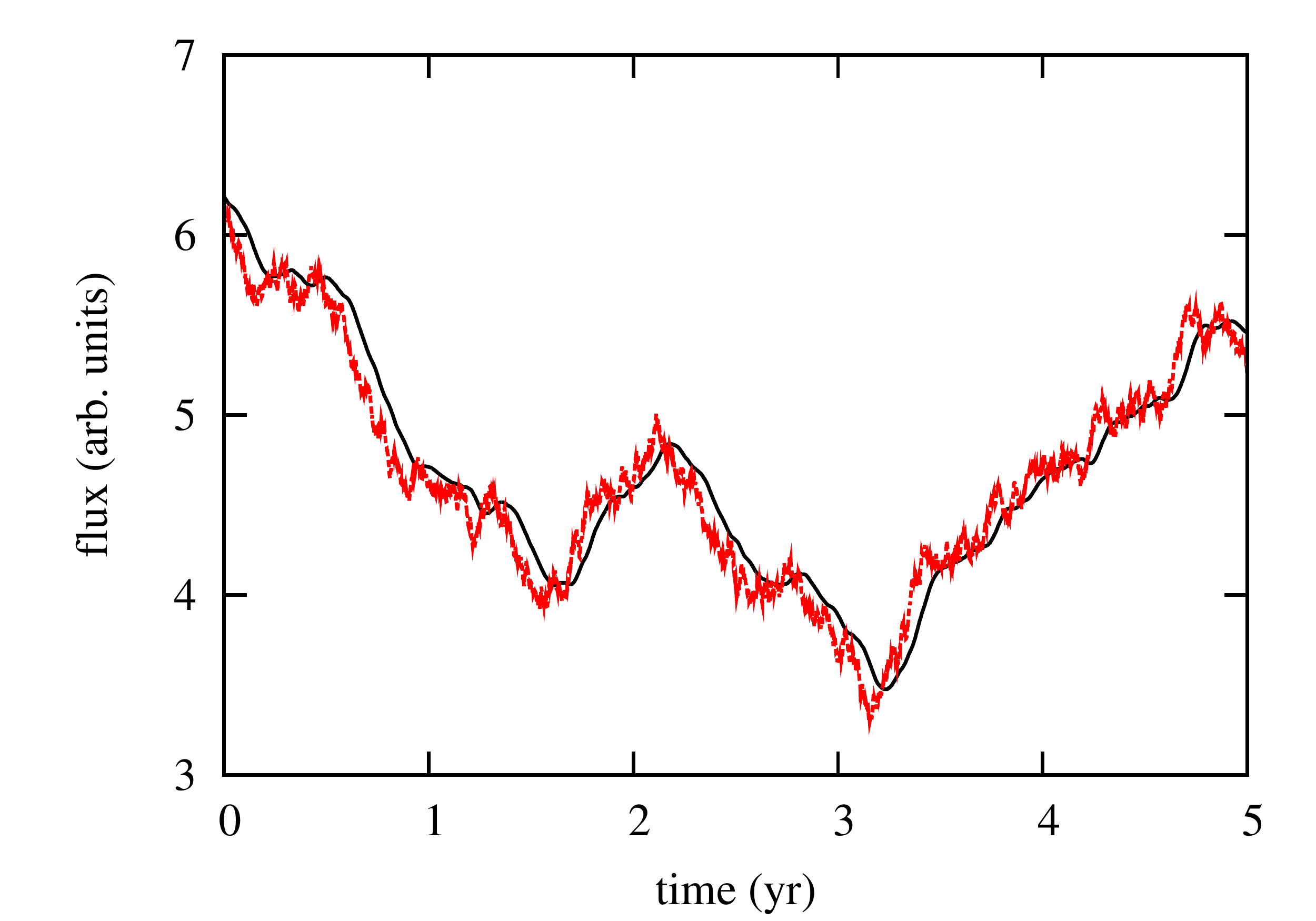}
  \caption{Total (red) and polarized (black) light curves obtained convolving the source term with the transfer function for a geometry including a dusty torus, a polar wind and an equatorial scattering ring with optical depth 1. The polarized flux has been rescaled by about one order of magnitude so that it can easily be compared to the total flux.}
  \label{Fig:lc}%
\end{figure}

\subsection{Cross-correlation analysis}

Figure \ref{Fig:lc} shows the total and polarized fluxes calculated using Eq. (\ref{eq:tfsrm}) for a geometry including a dusty torus, a polar wind and an equatorial scattering ring; the parameters are given in Table \ref{tab:param}. The effect of the scattering ring on the polarized radiation is clearly visible; it both smooths the short term fluctuations, and it introduces a delay that is visible to the eye, on time-scales as short as a few days.

We then performed a cross-correlation analysis of the total and polarized flux, after having removed the long term trend by subtracting the linear component of the light curve, defined as $L(0)+(L(T)-L(0))t/T$ where $T$ is the observation duration. The cross-correlation has a maximum for a positive time lag, but the position of the maximum varies somewhat for different source light curves. We thus generated, for each configuration, 100 different light cures, which enables us to estimate the mean delay and its standard deviation. Subtracting the long term behaviour of the light curve turns out to be essential for reducing the standard deviation of the measured time lag; the accuracy of the determination of this delay is improved by a factor larger than 10. This is because, for the time-scales we consider here, the mean and standard deviation are difficult to estimate because of the importance of the low frequency component in the light curves \citep[see e.g.][for a discussion of impact of the limits of the cross-correlation analysis in the case of short time series]{vw01}. Increasing the integration time will not solve the problem, because of the $\nu^{-2}$ frequency dependence of the signal in the case of a random walk.

We also stress that our modelled light curves do not include statistical and instrumental noise, nor gaps in data collection. These will introduce observational errors, and possibly biases in the measured lags. For lags of the order of a few days to a few weeks, the fact that sources are observable only at night during a fraction of a year will not be a limitation, but the availability of telescope time might be a problem. In any case, real observations will be analysed using more sophisticated methods, such as the discrete correlation function (DCF) or z-transformed DCF \citep[ZDCF,][]{a13}. Such methods also enable to estimate the observational error on the lag. We use here a simplified method, because we are not limited by observational constraints, and we note that, by construction, the DCF or the ZCDF method will provide the same results as ours on the time-lag for a continuous, large observational time.
\begin{table*}
\caption{Time-lag (in days) between the polarized and total flux for different configurations}
\label{tab:ccf}
\centering
\begin{tabular}{l c c c  }     
\hline  \hline \\ [-1.8ex]
component                                & V (5500 \AA)      & B (4500 \AA)      & UV (2500 \AA)     \\
\hline \\ [-1.8ex]
BLR                                      & $26.2 \pm 0.9$  & $26.4 \pm 0.9$  & $26.2 \pm 0.8$  \\
Uniform dusty torus                      & $90.7 \pm 2.4$  & $84.0 \pm 2.0$  & $68.5 \pm 1.8$  \\
Clumpy dusty torus                       & $686 \pm 24$    & $695  \pm 13$   & $ 678 \pm 17$   \\
Polar wind                               & $11500 \pm 850$& $11500 \pm 730$ & $11700 \pm 810$\\
Uniform dusty torus + BLR                & $29.1 \pm 1.0$  & $29.5 \pm 1.0$  & $28.3 \pm 1.2$  \\
Uniform dusty torus + polar wind         & $96.0 \pm 3.9$  & $89.7 \pm 3.0$  & $73.2 \pm 2.6$  \\
Uniform dusty torus + polar wind + BLR   & $29.6 \pm 1.2$  & $29.0 \pm 1.1$  & $28.5 \pm 0.8$  \\
Uniform flared disk + BLR                & $28.5 \pm 1.0$  & $28.3 \pm 1.3$  & $ 27.9 \pm 0.7$ \\
Uniform flared disk + polar wind         & $97.5 \pm 4.9$  & $92.8 \pm 5.0$  & $65.8 \pm 2.5$  \\
Uniform flared disk + polar wind + BLR   & $29.1 \pm 1.1$  & $28.5 \pm 1.2$  & $28.2 \pm 0.8$  \\
Clumpy dusty torus + BLR                 & $26.6 \pm 0.7$  & $25.9 \pm 0.8$  & $26.3 \pm 0.9$  \\
Clumpy dusty torus + polar wind          & $8260 \pm 1200$ & $8600 \pm 1200$ & $8750 \pm 1300$ \\
Clumpy dusty torus + polar wind + BLR    & $27.1 \pm 0.9$  & $27.3 \pm 0.9$  & $ 27.6 \pm 0.8$ \\
\hline
\end{tabular}
\end{table*}

\begin{table}
\caption{Time-lag (in days) between the polarized and total flux for different configurations}
\label{tab:tau_blr}
\centering
\begin{tabular}{l c c c  }     
\hline  \hline \\ [-1.8ex]
$\tau_{BLR}$                               &      lag \\
\hline \\ [-1.8ex]
1  &  $26.2 \pm 0.9$ \\
3  &  $16.2 \pm 0.5$ \\
10 &  $7.3  \pm 0.3$ \\
\hline
\end{tabular}
\end{table}

\section{Results} \label{sec:results}
Table \ref{tab:ccf} gives the time lag in days between the polarized and total flux, for various configurations, and for three observation wavelengths: 5500~\AA, corresponding to the V band, 4500~\AA{} for the B band, and 2500~\AA{} in the ultraviolet. As can be seen, the lags we find are not very different from the ones obtained when searching for the maximum of $\tau \Psi(\tau)$, as discussed in the previous section, with the exception of the polar wind case. In the latter case, the cross-correlation process picks up the first peak in $\tau \Psi(\tau)$. As expected, the delay obtained when considering the BLR and the polar wind alone do not depend on the observation wavelength, since Thomson scattering is a wavelength independent process. The uniform dusty environments (flared disk or torus) generates wavelength-dependent delays, as expected, but time-lags in clumpy environments do not depend on wavelength. This occurs because the depth at which photons can penetrate depends on the distribution of clumps and is thus wavelength independent, while the absorption process itself still depends on wavelength, but because the clumps are optically thick (the optical depth of each cloud is 50), occurs only in the skin of the clumps.

Our major result is that, whatever the configuration, the time-lag is determined by the presence or not of the BLR. When present, it essentially sets the lag to about 25 days; this lag is independent of the observation wavelength. It corresponds approximately to the the outer radius of the BLR, thereby validating the method of \citet{ggm12} to constrain the physical characteristics of the BLR; as mentioned in the introduction, we do not intend to reproduce the observed delay in one particular object, but rather to test the methodology.

It is interesting to compare these results to the average time lag we estimated in Paper I, and defined in Eq. \ref{eq:tau_av}. This average delay is significantly longer than the time lag found when calculating the peak of the cross-correlation function. In the case of a uniform dusty torus plus polar winds and BLR, this average time is of the order of 1.7~years, much longer than the 25 days found here.

One can also estimate the time lag between the polarized and total flux for different optical depths of the BLR. The results are given in Table \ref{tab:tau_blr}. As expected, the higher the optical depth, the smaller the lag, because for high optical depths, photons cannot penetrate deep in the equatorial ring and are scattered in regions closer to the ring inner edge. For reasonable values of the optical depths of the BLR (of the order of 1), the lag is expected to be of order of 25 days.

The accuracy to which the time lag can be measured depends indeed on the observing time. The values presented in Table \ref{tab:ccf} are calculated using very long baselines, much longer than what would be realistic: we use a duration of 80 years for the  uniform or clumpy torus plus polar wind and equatorial ring. For those long observing times, the uncertainty on our results partly arises from the discretization noise, as our time resolution is 0.6~d. For observations lasting for five years, this lag is still measurable with an accuracy of 7 days, and it remains detectable for observing times as short as 1 year. One must stress that the possibility of determining the time-lag with a reasonable degree of confidence strongly depends on the the statistics of the light fluctuations that are strongly source dependent. For one given source, the ability to determine a time lag also depends on the specific realization of the light curve noise, as can be anticipated when having a closer look at Fig. \ref{Fig:lc}. The most favourable situation occurs when a peak (or a minimum) whose width is slightly larger than the expected time-lag is detected during one observation, in which case the total observing time need not be longer than a few months. We finally note that the noise properties that we have chosen here -- random walk generating a $\nu^{-2}$ power spectrum -- is not favourable at all, because of the low frequencies, making it difficult (formally impossible) to determine the mean and standard deviation of the light curve.

We have checked that the time-lags we obtain do not depend sensitively on the assumption made for describing the time variability of the central source. Using a value of $\beta$ in Eq. \ref{eq:sterm} ten times larger or smaller than our reference value of 0.07 leads to delays for the clumpy dusty torus (CDT) + polar wind (PW) + BLR model that are within our uncertainties. Similarly, we considered a damped random walk model with a characteristic damping time of 20, 200 and 2000 days; we obtained for the CDT+PW+BLR models lags of $22.3 \pm 0.3$, $23.0 \pm 0.3$ and $24.2 \pm 1.8$ days respectively, in good agreement with the value of $26.2$ days obtained in the pure random walk case. This should not come as a surprise, given our interpretation of the time-lag as the maximum of $\tau \Psi(\tau)$.

\section{Conclusion}
We have shown that determining the delay between the polarized and total flux using cross correlation techniques enables one to constrain the size of the BLR, as first suggested by \citet{ggm12}. Matter located at larger distance, as e.g. in the dusty torus or in polar winds, do not contribute to the time-lag as determined using this technique, although it does increase the average propagation time of a photon. This happens because this external matter does not significantly modify the peak of the transfer function, but introduces wings that can be significant enough to change by a large factor the mean travel time. This difference between the time-lag determined by cross-correlation techniques and the average propagation time delay is also due to the fact that short term fluctuations in the source light curve may prevent one to find correlations on very long time scales. The only case where the outer environment may affect the measured time-lag is when the time-lag generated by these regions is commensurable with the lag due to the BLR; in our case when the dusty torus is not clumpy. In any case, this effect is moderate.

We have also shown that this time-lag is measurable as long as the observing time span covered by observations is longer than a few years, and possibly shorter, depending on the statistical properties of the light fluctuations of a given source or on the actual realisation of the noise.

One should note, however, that the time-lag between the polarized and unpolarized radiation does not univocally determine the BLR size, because parameters such as in particular the AGN viewing angle and the BLR optical depth must be known a priori. These are reasonably constrained -- the viewing angle must not be large for a type 1 AGN, and the BLR optical depth cannot be significantly larger than unity -- but their uncertainty can be significant \citep[see e.g.][for a discussion on the determination of the inclination angle]{m16}. We have also considered a homogeneous scattering ring; clumps are certainly present in the BLR \citep{g09}, but, because the BLR optical depth cannot be large, the presence of clumps is not as dramatic as it is for instance in the case of a dusty torus.

Better constraints might be obtained by considering polarized lines, as discussed in e.g. \citet{pas18} and \citet{sgp18}, since the width of the line directly measures the radial velocity of the material, and is thus connected to the position of the emitting material in the BLR. We leave this for a future work.

As a final remark, it is worth mentioning that this method can be used for low luminosity sources, such as for example the Galactic centre. The immediate vicinity of Sgr~A* contains several gaseous structures \citep{ccm20,phz20} whose nature is still uncertain; they are unresolved and show, at least for the so-called G objects, emission properties of gas clouds but their dynamical properties are typical of stellar objects. One of them, G2, is intrinsically linearly polarized \citep{sez16} in the infrared ($K_{\rm s}$ band), with a polarized fraction larger than 20\%, and  a varying polarization angle as it approaches the position of Sgr A*. If the infrared luminosity of these clouds results, at least for a sizeable fraction, from the reprocessing of Sgr A* luminosity, one would expect a delayed response of the polarized radiation of these sources with respect to Sgr A*.

\begin{acknowledgements}
We are grateful to D. Savic and M. Gaskell for helpful discussions. We are grateful to the referee for helpful comments and suggestions, in particular concerning the applicability of our method to the Galactic centre. This research has been supported by the French Programme National des Hautes Energies (PNHE). FM is grateful for CNES funding under the post-doctoral grant ``Probing the geometry and physics of active galactic nuclei with ultraviolet and X-ray polarized radiative transfer''. APRL acknowledge support from the CONICYT BECAS Chile grant no. 72150573.
\end{acknowledgements}

\bibliographystyle{aa}
\bibliography{biblio}

\begin{thebibliography}{45}
\expandafter\ifx\csname natexlab\endcsname\relax\def\natexlab#1{#1}\fi

\bibitem[{{Alexander}(2013)}]{a13}
{Alexander}, T. 2013, arXiv e-prints, arXiv:1302.1508

\bibitem[{{Antonucci}(1993)}]{a93}
{Antonucci}, R. 1993, \araa, 31, 473

\bibitem[{{Antonucci}(1984)}]{a84b}
{Antonucci}, R.~R.~J. 1984, \apj, 278, 499

\bibitem[{{Antonucci} \& {Miller}(1985)}]{am85}
{Antonucci}, R.~R.~J. \& {Miller}, J.~S. 1985, \apj, 297, 621

\bibitem[{{Aranzana} {et~al.}(2018){Aranzana}, {K{\"o}rding}, {Uttley},
  {Scaringi}, \& {Bloemen}}]{aku18}
{Aranzana}, E., {K{\"o}rding}, E., {Uttley}, P., {Scaringi}, S., \& {Bloemen},
  S. 2018, \mnras, 476, 2501

\bibitem[{{Bentz} {et~al.}(2013){Bentz}, {Denney}, {Grier}, {Barth},
  {Peterson}, {Vestergaard}, {Bennert}, {Canalizo}, {De Rosa}, {Filippenko},
  {Gates}, {Greene}, {Li}, {Malkan}, {Pogge}, {Stern}, {Treu}, \&
  {Woo}}]{bdg13}
{Bentz}, M.~C., {Denney}, K.~D., {Grier}, C.~J., {et~al.} 2013, \apj, 767, 149

\bibitem[{{Blandford} \& {McKee}(1982)}]{bm88}
{Blandford}, R.~D. \& {McKee}, C.~F. 1982, \apj, 255, 419

\bibitem[{{Chandrasekhar}(1960)}]{c60}
{Chandrasekhar}, S. 1960, {Radiative transfer} (Dover Publications, New
  York,USA)

\bibitem[{{Ciurlo} {et~al.}(2020){Ciurlo}, {Campbell}, {Morris}, {Do}, {Ghez},
  {Hees}, {Sitarski}, {Kosmo O'Neil}, {Chu}, {Martinez}, {Naoz}, \&
  {Stephan}}]{ccm20}
{Ciurlo}, A., {Campbell}, R.~D., {Morris}, M.~R., {et~al.} 2020, \nat, 577, 337

\bibitem[{{Czerny} \& {Hryniewicz}(2011)}]{ch11}
{Czerny}, B. \& {Hryniewicz}, K. 2011, \aap, 525, L8

\bibitem[{{Czerny} {et~al.}(2017){Czerny}, {Li}, {Hryniewicz}, {Panda},
  {Wildy}, {Sniegowska}, {Wang}, {Sredzinska}, \& {Karas}}]{clh17}
{Czerny}, B., {Li}, Y.-R., {Hryniewicz}, K., {et~al.} 2017, \apj, 846, 154

\bibitem[{{Czerny} {et~al.}(2019){Czerny}, {Olejak}, {Ra{\l}owski},
  {Koz{\l}owski}, {Loli Martinez Aldama}, {Zajacek}, {Pych}, {Hryniewicz},
  {Pietrzy{\'n}ski}, {Sobrino Figaredo}, {Haas}, {{\'S}redzi{\'n}ska}, {Krupa},
  {Kurcz}, {Udalski}, {Gorski}, {Karas}, {Panda}, {Sniegowska}, {Naddaf},
  {Bilicki}, \& {Sarna}}]{cor19}
{Czerny}, B., {Olejak}, A., {Ra{\l}owski}, M., {et~al.} 2019, \apj, 880, 46

\bibitem[{{Event Horizon Telescope Collaboration} {et~al.}(2019){Event Horizon
  Telescope Collaboration}, {Akiyama}, {Alberdi}, {Alef}, {Asada}, {Azulay},
  {Baczko}, {Ball}, {Balokovi{\'c}}, {Barrett}, {Bintley}, {Blackburn},
  {Boland}, {Bouman}, {Bower}, {Bremer}, {Brinkerink}, {Brissenden}, {Britzen},
  {Broderick}, {Broguiere}, {Bronzwaer}, {Byun}, {Carlstrom}, {Chael}, {Chan},
  {Chatterjee}, {Chatterjee}, {Chen}, {Chen}, {Cho}, {Christian}, {Conway},
  {Cordes}, {Crew}, {Cui}, {Davelaar}, {De Laurentis}, {Deane}, {Dempsey},
  {Desvignes}, {Dexter}, {Doeleman}, {Eatough}, {Falcke}, {Fish}, {Fomalont},
  {Fraga-Encinas}, {Freeman}, {Friberg}, {Fromm}, {G{\'o}mez}, {Galison},
  {Gammie}, {Garc{\'\i}a}, {Gentaz}, {Georgiev}, {Goddi}, {Gold}, {Gu},
  {Gurwell}, {Hada}, {Hecht}, {Hesper}, {Ho}, {Ho}, {Honma}, {Huang}, {Huang},
  {Hughes}, {Ikeda}, {Inoue}, {Issaoun}, {James}, {Jannuzi}, {Janssen},
  {Jeter}, {Jiang}, {Johnson}, {Jorstad}, {Jung}, {Karami}, {Karuppusamy},
  {Kawashima}, {Keating}, {Kettenis}, {Kim}, {Kim}, {Kim}, {Kino}, {Koay},
  {Koch}, {Koyama}, {Kramer}, {Kramer}, {Krichbaum}, {Kuo}, {Lauer}, {Lee},
  {Li}, {Li}, {Lindqvist}, {Liu}, {Liuzzo}, {Lo}, {Lobanov}, {Loinard},
  {Lonsdale}, {Lu}, {MacDonald}, {Mao}, {Markoff}, {Marrone}, {Marscher},
  {Mart{\'\i}-Vidal}, {Matsushita}, {Matthews}, {Medeiros}, {Menten}, {Mizuno},
  {Mizuno}, {Moran}, {Moriyama}, {Moscibrodzka}, {M{\"u}ller}, {Nagai},
  {Nagar}, {Nakamura}, {Narayan}, {Narayanan}, {Natarajan}, {Neri}, {Ni},
  {Noutsos}, {Okino}, {Olivares}, {Ortiz-Le{\'o}n}, {Oyama}, {{\"O}zel},
  {Palumbo}, {Patel}, {Pen}, {Pesce}, {Pi{\'e}tu}, {Plambeck}, {PopStefanija},
  {Porth}, {Prather}, {Preciado-L{\'o}pez}, {Psaltis}, {Pu}, {Ramakrishnan},
  {Rao}, {Rawlings}, {Raymond}, {Rezzolla}, {Ripperda}, {Roelofs}, {Rogers},
  {Ros}, {Rose}, {Roshanineshat}, {Rottmann}, {Roy}, {Ruszczyk}, {Ryan},
  {Rygl}, {S{\'a}nchez}, {S{\'a}nchez-Arguelles}, {Sasada}, {Savolainen},
  {Schloerb}, {Schuster}, {Shao}, {Shen}, {Small}, {Sohn}, {SooHoo}, {Tazaki},
  {Tiede}, {Tilanus}, {Titus}, {Toma}, {Torne}, {Trent}, {Trippe}, {Tsuda},
  {van Bemmel}, {van Langevelde}, {van Rossum}, {Wagner}, {Wardle},
  {Weintroub}, {Wex}, {Wharton}, {Wielgus}, {Wong}, {Wu}, {Young}, {Young},
  {Younsi}, {Yuan}, {Yuan}, {Zensus}, {Zhao}, {Zhao}, {Zhu}, {Algaba},
  {Allardi}, {Amestica}, {Anczarski}, {Bach}, {Baganoff}, {Beaudoin}, {Benson},
  {Berthold}, {Blanchard}, {Blundell}, {Bustamente}, {Cappallo},
  {Castillo-Dom{\'\i}nguez}, {Chang}, {Chang}, {Chang}, {Chen}, {Chilson},
  {Chuter}, {C{\'o}rdova Rosado}, {Coulson}, {Crawford}, {Crowley}, {David},
  {Derome}, {Dexter}, {Dornbusch}, {Dudevoir}, {Dzib}, {Eckart}, {Eckert},
  {Erickson}, {Everett}, {Faber}, {Farah}, {Fath}, {Folkers}, {Forbes},
  {Freund}, {G{\'o}mez-Ruiz}, {Gale}, {Gao}, {Geertsema}, {Graham}, {Greer},
  {Grosslein}, {Gueth}, {Haggard}, {Halverson}, {Han}, {Han}, {Hao},
  {Hasegawa}, {Henning}, {Hern{\'a}ndez-G{\'o}mez}, {Herrero-Illana},
  {Heyminck}, {Hirota}, {Hoge}, {Huang}, {Impellizzeri}, {Jiang}, {Kamble},
  {Keisler}, {Kimura}, {Kono}, {Kubo}, {Kuroda}, {Lacasse}, {Laing}, {Leitch},
  {Li}, {Lin}, {Liu}, {Liu}, {Lu}, {Marson}, {Martin-Cocher}, {Massingill},
  {Matulonis}, {McColl}, {McWhirter}, {Messias}, {Meyer-Zhao}, {Michalik},
  {Monta{\~n}a}, {Montgomerie}, {Mora-Klein}, {Muders}, {Nadolski}, {Navarro},
  {Neilsen}, {Nguyen}, {Nishioka}, {Norton}, {Nowak}, {Nystrom}, {Ogawa},
  {Oshiro}, {Oyama}, {Parsons}, {Paine}, {Pe{\~n}alver}, {Phillips}, {Poirier},
  {Pradel}, {Primiani}, {Raffin}, {Rahlin}, {Reiland}, {Risacher}, {Ruiz},
  {S{\'a}ez-Mada{\'\i}n}, {Sassella}, {Schellart}, {Shaw}, {Silva}, {Shiokawa},
  {Smith}, {Snow}, {Souccar}, {Sousa}, {Sridharan}, {Srinivasan}, {Stahm},
  {Stark}, {Story}, {Timmer}, {Vertatschitsch}, {Walther}, {Wei}, {Whitehorn},
  {Whitney}, {Woody}, {Wouterloot}, {Wright}, {Yamaguchi}, {Yu}, {Zeballos},
  {Zhang}, \& {Ziurys}}]{eht19}
{Event Horizon Telescope Collaboration}, {Akiyama}, K., {Alberdi}, A., {et~al.}
  2019, \apjl, 875, L1

\bibitem[{{Garc{\'\i}a-Burillo} {et~al.}(2016){Garc{\'\i}a-Burillo}, {Combes},
  {Ramos Almeida}, {Usero}, {Krips}, {Alonso-Herrero}, {Aalto}, {Casasola},
  {Hunt}, {Mart{\'\i}n}, {Viti}, {Colina}, {Costagliola}, {Eckart}, {Fuente},
  {Henkel}, {M{\'a}rquez}, {Neri}, {Schinnerer}, {Tacconi}, \& {van der
  Werf}}]{gcr16}
{Garc{\'\i}a-Burillo}, S., {Combes}, F., {Ramos Almeida}, C., {et~al.} 2016,
  \apjl, 823, L12

\bibitem[{{Gaskell}(2009)}]{g09}
{Gaskell}, C.~M. 2009, \nar, 53, 140

\bibitem[{{Gaskell} {et~al.}(2012){Gaskell}, {Goosmann}, {Merkulova},
  {Shakhovskoy}, \& {Shoji}}]{ggm12}
{Gaskell}, C.~M., {Goosmann}, R.~W., {Merkulova}, N.~I., {Shakhovskoy}, N.~M.,
  \& {Shoji}, M. 2012, \apj, 749, 148

\bibitem[{{Goosmann} \& {Gaskell}(2007)}]{gg07}
{Goosmann}, R.~W. \& {Gaskell}, C.~M. 2007, \aap, 465, 129

\bibitem[{{Gravity Collaboration} {et~al.}(2018{\natexlab{a}}){Gravity
  Collaboration}, {Abuter}, {Amorim}, {Baub{\"o}ck}, {Berger}, {Bonnet},
  {Brandner}, {Cl{\'e}net}, {Coud{\'e} Du Foresto}, {de Zeeuw}, {Deen},
  {Dexter}, {Duvert}, {Eckart}, {Eisenhauer}, {F{\"o}rster Schreiber},
  {Garcia}, {Gao}, {Gendron}, {Genzel}, {Gillessen}, {Guajardo}, {Habibi},
  {Haubois}, {Henning}, {Hippler}, {Horrobin}, {Huber}, {Jim{\'e}nez-Rosales},
  {Jocou}, {Kervella}, {Lacour}, {Lapeyr{\`e}re}, {Lazareff}, {Le Bouquin},
  {L{\'e}na}, {Lippa}, {Ott}, {Panduro}, {Paumard}, {Perraut}, {Perrin},
  {Pfuhl}, {Plewa}, {Rabien}, {Rodr{\'{\i}}guez-Coira}, {Rousset}, {Sternberg},
  {Straub}, {Straubmeier}, {Sturm}, {Tacconi}, {Vincent}, {von Fellenberg},
  {Waisberg}, {Widmann}, {Wieprecht}, {Wiezorrek}, {Woillez}, \&
  {Yazici}}]{gaa18}
{Gravity Collaboration}, {Abuter}, R., {Amorim}, A., {et~al.}
  2018{\natexlab{a}}, \aap, 618, L10

\bibitem[{{GRAVITY Collaboration} {et~al.}(2019){GRAVITY Collaboration},
  {Dexter}, {Shangguan}, {H{\"o}nig}, {Kishimoto}, {Lutz}, {Netzer}, {Davies},
  {Sturm}, {Pfuhl}, {Amorim}, {Baub{\"o}ck}, {Brandner}, {Cl{\'e}net}, {de
  Zeeuw}, {Eckart}, {Eisenhauer}, {F{\"o}rster Schreiber}, {Gao}, {Garcia},
  {Genzel}, {Gillessen}, {Gratadour}, {Jim{\'e}nez-Rosales}, {Lacour},
  {Millour}, {Ott}, {Paumard}, {Perraut}, {Perrin}, {Peterson}, {Petrucci},
  {Prieto}, {Rouan}, {Schartmann}, {Shimizu}, {Sternberg}, {Straub},
  {Straubmeier}, {Tacconi}, {Tristram}, {Vermot}, {Waisberg}, {Widmann}, \&
  {Woillez}}]{gds19}
{GRAVITY Collaboration}, {Dexter}, J., {Shangguan}, J., {et~al.} 2019, arXiv
  e-prints, arXiv:1910.00593

\bibitem[{{Gravity Collaboration} {et~al.}(2018{\natexlab{b}}){Gravity
  Collaboration}, {Sturm}, {Dexter}, {Pfuhl}, {Stock}, {Davies}, {Lutz},
  {Cl{\'e}net}, {Eckart}, {Eisenhauer}, {Genzel}, {Gratadour}, {H{\"o}nig},
  {Kishimoto}, {Lacour}, {Millour}, {Netzer}, {Perrin}, {Peterson}, {Petrucci},
  {Rouan}, {Waisberg}, {Woillez}, {Amorim}, {Brandner}, {F{\"o}rster
  Schreiber}, {Garcia}, {Gillessen}, {Ott}, {Paumard}, {Perraut},
  {Scheithauer}, {Straubmeier}, {Tacconi}, \& {Widmann}}]{gsd18}
{Gravity Collaboration}, {Sturm}, E., {Dexter}, J., {et~al.}
  2018{\natexlab{b}}, \nat, 563, 657

\bibitem[{{Jaffe} {et~al.}(2004){Jaffe}, {Meisenheimer}, {R{\"o}ttgering},
  {Leinert}, \& {Richichi}}]{jmr04}
{Jaffe}, W., {Meisenheimer}, K., {R{\"o}ttgering}, H., {Leinert}, C., \&
  {Richichi}, A. 2004, in IAU Symposium, Vol. 222, The Interplay Among Black
  Holes, Stars and ISM in Galactic Nuclei, ed. T.~{Storchi-Bergmann}, L.~C.
  {Ho}, \& H.~R. {Schmitt}, 37--39

\bibitem[{{Kelly} {et~al.}(2009){Kelly}, {Bechtold}, \&
  {Siemiginowska}}]{kbs09}
{Kelly}, B.~C., {Bechtold}, J., \& {Siemiginowska}, A. 2009, \apj, 698, 895

\bibitem[{{Kelly} {et~al.}(2014){Kelly}, {Becker}, {Sobolewska},
  {Siemiginowska}, \& {Uttley}}]{kbs14}
{Kelly}, B.~C., {Becker}, A.~C., {Sobolewska}, M., {Siemiginowska}, A., \&
  {Uttley}, P. 2014, \apj, 788, 33

\bibitem[{{Lira} {et~al.}(2018){Lira}, {Kaspi}, {Netzer}, {Botti}, {Morrell},
  {Mej{\'\i}a-Restrepo}, {S{\'a}nchez-S{\'a}ez}, {Mart{\'\i}nez-Palomera}, \&
  {L{\'o}pez}}]{lkn18}
{Lira}, P., {Kaspi}, S., {Netzer}, H., {et~al.} 2018, \apj, 865, 56

\bibitem[{{MacLeod} {et~al.}(2010){MacLeod}, {Ivezi{\'c}}, {Kochanek},
  {Koz{\l}owski}, {Kelly}, {Bullock}, {Kimball}, {Sesar}, {Westman}, {Brooks},
  {Gibson}, {Becker}, \& {de Vries}}]{mik10}
{MacLeod}, C.~L., {Ivezi{\'c}}, {\v Z}., {Kochanek}, C.~S., {et~al.} 2010,
  \apj, 721, 1014

\bibitem[{{Marin}(2016)}]{m16}
{Marin}, F. 2016, \mnras, 460, 3679

\bibitem[{{Marin}(2018)}]{m18}
{Marin}, F. 2018, \aap, 615, A171

\bibitem[{{Marin} \& {Goosmann}(2013)}]{mg13}
{Marin}, F. \& {Goosmann}, R.~W. 2013, \mnras, 436, 2522

\bibitem[{{Marin} {et~al.}(2015){Marin}, {Goosmann}, \& {Gaskell}}]{mgg15}
{Marin}, F., {Goosmann}, R.~W., \& {Gaskell}, C.~M. 2015, \aap, 577, A66

\bibitem[{{Marin} {et~al.}(2012){Marin}, {Goosmann}, {Gaskell}, {Porquet}, \&
  {Dov{\v c}iak}}]{mgg12}
{Marin}, F., {Goosmann}, R.~W., {Gaskell}, C.~M., {Porquet}, D., \& {Dov{\v
  c}iak}, M. 2012, \aap, 548, A121

\bibitem[{{Miller} \& {Antonucci}(1983)}]{ma83}
{Miller}, J.~S. \& {Antonucci}, R.~R.~J. 1983, \apjl, 271, L7

\bibitem[{{Pei{\ss}ker} {et~al.}(2020){Pei{\ss}ker}, {Hosseini},
  {Zaja{\v{c}}ek}, {Eckart}, {Saalfeld}, {Valencia-S.}, {Parsa}, \&
  {Karas}}]{phz20}
{Pei{\ss}ker}, F., {Hosseini}, S.~E., {Zaja{\v{c}}ek}, M., {et~al.} 2020, \aap,
  634, A35

\bibitem[{{Peterson}(1993)}]{p93}
{Peterson}, B.~M. 1993, \pasp, 105, 247

\bibitem[{{Popovic} {et~al.}(2018){Popovic}, {Afanasiev}, \& {Savic}}]{pas18}
{Popovic}, L.~C., {Afanasiev}, V.~L., \& {Savic}, D. 2018, arXiv e-prints,
  arXiv:1807.00177

\bibitem[{{Raban} {et~al.}(2009){Raban}, {Jaffe}, {R{\"o}ttgering},
  {Meisenheimer}, \& {Tristram}}]{rjr09}
{Raban}, D., {Jaffe}, W., {R{\"o}ttgering}, H., {Meisenheimer}, K., \&
  {Tristram}, K. R.~W. 2009, \mnras, 394, 1325

\bibitem[{{Rojas Lobos} {et~al.}(2018){Rojas Lobos}, {Goosmann}, {Marin}, \&
  {Savi{\'c}}}]{rgm18}
{Rojas Lobos}, P.~A., {Goosmann}, R.~W., {Marin}, F., \& {Savi{\'c}}, D. 2018,
  \aap, 611, A39

\bibitem[{{Savi{\'c}} {et~al.}(2018){Savi{\'c}}, {Goosmann}, {Popovi{\'c}},
  {Marin}, \& {Afanasiev}}]{sgp18}
{Savi{\'c}}, D., {Goosmann}, R., {Popovi{\'c}}, L.~{\v{C}}., {Marin}, F., \&
  {Afanasiev}, V.~L. 2018, \aap, 614, A120

\bibitem[{{Shahzamanian} {et~al.}(2016){Shahzamanian}, {Eckart},
  {Zaja{\v{c}}ek}, {Valencia-S.}, {Sabha}, {Moser}, {Parsa}, {Peissker}, \&
  {Straubmeier}}]{sez16}
{Shahzamanian}, B., {Eckart}, A., {Zaja{\v{c}}ek}, M., {et~al.} 2016, \aap,
  593, A131

\bibitem[{{Shoji} {et~al.}(2005){Shoji}, {Gaskell}, \& {Goosmann}}]{sgg05}
{Shoji}, M., {Gaskell}, C.~M., \& {Goosmann}, R.~W. 2005, in Bulletin of the
  American Astronomical Society, Vol.~37, American Astronomical Society Meeting
  Abstracts, 1420

\bibitem[{{Smith} {et~al.}(2004){Smith}, {Robinson}, {Alexander}, {Young},
  {Axon}, \& {Corbett}}]{sra04}
{Smith}, J.~E., {Robinson}, A., {Alexander}, D.~M., {et~al.} 2004, \mnras, 350,
  140

\bibitem[{{Timmer} \& {Koenig}(1995)}]{tk95}
{Timmer}, J. \& {Koenig}, M. 1995, \aap, 300, 707

\bibitem[{{Tombesi} {et~al.}(2013){Tombesi}, {Cappi}, {Reeves}, {Nemmen},
  {Braito}, {Gaspari}, \& {Reynolds}}]{tcr13}
{Tombesi}, F., {Cappi}, M., {Reeves}, J.~N., {et~al.} 2013, \mnras, 430, 1102

\bibitem[{{Vio} \& {Wamsteker}(2001)}]{vw01}
{Vio}, R. \& {Wamsteker}, W. 2001, \pasp, 113, 86

\bibitem[{{Zu} {et~al.}(2013){Zu}, {Kochanek}, {Koz{\l}owski}, \&
  {Udalski}}]{zkk13}
{Zu}, Y., {Kochanek}, C.~S., {Koz{\l}owski}, S., \& {Udalski}, A. 2013, \apj,
  765, 106

\bibitem[{{Zu} {et~al.}(2011){Zu}, {Kochanek}, \& {Peterson}}]{zkp11}
{Zu}, Y., {Kochanek}, C.~S., \& {Peterson}, B.~M. 2011, \apj, 735, 80

\end{thebibliography}

\end{document}